\documentclass[twocolumn]{aastex63}

\newcommand{\kms}{\mathrm{km\,s^{-1}}}

\usepackage{amsmath}
\usepackage{comment}

\newcommand{\masuda}{2021ApJ...910L..17M}
\newcommand{\eb}{2022MNRAS.512.5620E}
\newcommand{\jaya}{2021MNRAS.504.2577J}
\defcitealias{2021ApJ...910L..17M}{MH21}
\defcitealias{2022MNRAS.512.5620E}{E22}

\newcommand{\teff}{T_\mathrm{eff}}
\newcommand{\logg}{\log g}
\newcommand{\mh}{\mathrm{[M/H]}}
\newcommand{\afe}{\mathrm{[\alpha/Fe]}}
\newcommand{\vsini}{v\sin i}
\newcommand{\unif}{\mathcal{U}}
\newcommand{\lunif}{\mathcal{LU}}

\newcommand{\mone}{0.46^{+0.12}_{-0.09}\,M_\odot}
\newcommand{\mtwo}{2.5\pm0.2\,M_\odot}

\newcommand{\moneeb}{0.44\pm0.06\,M_\odot}
\newcommand{\mtwoeb}{2.8\pm0.3\,M_\odot}

\newcommand{\inceb}{68\pm4\,\mathrm{deg}}

\shorttitle{Binary Mass from Tidal RVs}
\shortauthors{Tomoyoshi et al.}
\graphicspath{{./}{figures/}}

\begin{document}
\turnoffeditone

\title{ 
Weighing Single-lined Spectroscopic Binaries Using Tidal Effects on Radial Velocities:\\ The Case of V723 Monocerotis\footnote{This research is based on data collected at the Subaru Telescope, which is operated by the National Astronomical Observatory of Japan. We are honored and grateful for the opportunity of observing the Universe from Maunakea, which has the cultural, historical, and natural significance in Hawaii.}
}

\correspondingauthor{Kento Masuda}
\email{kmasuda@ess.sci.osaka-u.ac.jp}

\author[0009-0005-5192-9690]{Mio Tomoyoshi}

\author[0000-0003-1298-9699]{Kento Masuda}
\affiliation{Department of Earth and Space Science, Osaka University, Osaka 560-0043, Japan}

\author[0000-0003-3618-7535]{Teruyuki Hirano} 
\affiliation{Astrobiology Center, NINS, 2-21-1 Osawa, Mitaka, Tokyo 181-8588, Japan}
\affiliation{National Astronomical Observatory of Japan, NINS, 2-21-1 Osawa, Mitaka, Tokyo 181-8588, Japan}

\author[0000-0002-8607-358X]{Yui Kasagi}
\affiliation{Department of Space Astronomy and Astrophysics, ISAS/JAXA, 3-1-1, Yoshinodai, Sagamihara, Kanagawa, 252-5210 Japan}

\author[0000-0003-3309-9134]{Hajime Kawahara}
\affiliation{Department of Space Astronomy and Astrophysics, ISAS/JAXA, 3-1-1, Yoshinodai, Sagamihara, Kanagawa, 252-5210 Japan}
\affiliation{Department of Astronomy, Graduate School of Science, The University of Tokyo, 7-3-1 Hongo, Bunkyo-ku, Tokyo 113-0033, Japan}

\author[0000-0001-6181-3142]{Takayuki Kotani}
\affiliation{Astrobiology Center, NINS, 2-21-1 Osawa, Mitaka, Tokyo 181-8588, Japan}
\affiliation{National Astronomical Observatory of Japan, NINS, 2-21-1 Osawa, Mitaka, Tokyo 181-8588, Japan}
\affiliation{Department of Astronomical Science, The Graduate University for Advanced Studies, SOKENDAI, 2-21-1 Osawa, Mitaka, Tokyo 181-8588, Japan}

\author[0000-0002-9294-1793]{Tomoyuki Kudo}
\affiliation{Subaru Telescope, 650 North A’ohoku Place, Hilo, HI 96720, USA}

\author[0000-0002-6510-0681]{Motohide Tamura}
\affiliation{Department of Astronomy, Graduate School of Science, The University of Tokyo, 7-3-1 Hongo, Bunkyo-ku, Tokyo 113-0033, Japan}
\affiliation{Astrobiology Center, NINS, 2-21-1 Osawa, Mitaka, Tokyo 181-8588, Japan}
\affiliation{National Astronomical Observatory of Japan, NINS, 2-21-1 Osawa, Mitaka, Tokyo 181-8588, Japan}

\author[0000-0003-4018-2569]{S\'ebastien Vievard}
\affiliation{Astrobiology Center, NINS, 2-21-1 Osawa, Mitaka, Tokyo 181-8588, Japan}
\affiliation{Subaru Telescope, 650 North A’ohoku Place, Hilo, HI 96720, USA}




\begin{abstract}

In single-lined spectroscopic binaries (SB1s) where flux variations due to tidal deformation of the primary star (ellipsoidal variations, EVs) are detected, the binary mass can be determined by combining EVs with the primary’s radial velocity (RV) variations from orbital motion and information about the primary's radius. This method has been used for mass estimation in close binaries including X-ray systems, but it has been pointed out that contaminating light from sources other than the primary star could introduce systematic errors in the mass and inclination estimates. Here, we focus on the apparent RV variations caused by asymmetric distortion of the absorption lines of the tidally deformed primary star (tidal RV). Because this signal contains information equivalent to that from photometric EVs, it enables mass estimation of the binary system using only the primary star's absorption lines from high-resolution spectroscopic data, providing a potentially more robust approach against contaminating light. We apply the method to the binary system V723 Monocerotis, where both photometric EV and tidal RV signals are detected, and successfully determine the component masses using only the primary star's RVs and projected rotational velocity, without relying on absolute flux measurements or on stellar evolutionary models. The masses derived from the tidal RV model show a reasonable agreement with those obtained from EVs after carefully modeling the flux contamination from the secondary. This result demonstrates that tidal RVs provide a useful alternative means for mass estimation in SB1s. 

\end{abstract}

\keywords{Radial velocity(1332); Red giant stars(1372); Stellar mass black holes(1611); Tidal distortion(1697)}


\section{Introduction}\label{sec:intro}

Many important binary systems, including black hole binaries and exoplanetary systems, are observed as single-lined spectroscopic binaries (SB1s). 
In a tight SB1 that exhibits flux variations due to tidal deformation of the primary star\footnote{Throughout the paper, the ``primary'' refers to the brighter component of an SB1 for which radial velocities are measured. Thus the primary is not necessarily more massive than the secondary.} (ellipsoidal variations, EVs), the binary mass can be determined by combining EVs with the primary’s radial velocity (RV) variations due to orbital motion and information of the primary's radius (see Section~\ref{ssec:intro_principle}). This method has been used for mass estimation in close binaries including X-ray systems, but it has been pointed out that contaminating light from sources other than the primary star, 
if not properly taken into account, can introduce systematic errors in the mass and inclination estimates \citep[][see also Section~\ref{ssec:intro_v723mon}]{2012ApJ...757...36K}.

Here, we focus on spectroscopic effects of tidal deformation as an alternative means for binary mass estimation. The deformation of the primary star does not only change the apparent size of the star (that causes EVs), it also changes the relative area of the approaching and receding parts of the visible stellar surface in phase with the orbit. This results in asymmetric distortion of the Doppler-broadened absorption lines, and induces apparent RV variations \citep{1941PNAS...27..168S, 1959cbs..book.....K, 1976ApJ...203..182W}; see also Figure~1 of \citet{\masuda}. Because this ``tidal RV'' signal shares a common origin as EVs, it should in principle enable mass estimation of \edit1{an SB1} system using only the primary star's absorption lines from high-resolution spectroscopic data, providing a potentially more robust approach against flux contamination \edit1{that reduces the photometric EV amplitude} \citep{\masuda}. This effect on RVs has been detected in a handful of close binary systems involving evolved stars \citep{1989A&A...218..152H, 2008ApJ...681..562E}, but has not been used for quantitative mass measurements to our knowledge (see also Section~\ref{ssec:intro_tidalrv}).

In this paper, we apply a quantitative model of tidal RV signals developed in \citet{\masuda} to the binary system V723 Monocerotis to test the effectiveness of this method. The system consists of a red giant primary and a more compact, hotter subgiant secondary \citep{\eb}, where significant tidal deformation of the primary has enabled detection of both EVs and tidal RV signals \citep{\jaya, \masuda}; 
see also Section~\ref{ssec:intro_v723mon}.
We show that the component masses can be derived using only the primary star's absorption lines, without relying on the absolute flux information nor on stellar evolutionary models (Sections~\ref{sec:spectrum} and \ref{sec:tidalrv}). We also find that the result is reasonably accurate even in the presence of the flux contamination from the secondary star, based on the comparison with the EV-based mass estimate that takes into account the secondary flux explicitly (Section~\ref{sec:discussion}).
This result demonstrates that modeling of tidal RV signals is useful for secure mass estimation in tidally deformed SB1s of particular interest.

\subsection{Relations between the Masses and the Observables}\label{ssec:intro_principle}

In SB1s, the information on the physical dimensions of the system from the RV variations due to orbital motion is summarized in the binary mass function: 
\begin{align}
\label{eq:fm}
    {(M_2\sin i)^3 \over (M_1+M_2)^2} = {PK_1^3(1-e^2)^{3/2} \over 2\pi G},
\end{align}
where $M_1$ and $M_2$ are the masses of the primary and secondary stars, $K_1$ is the RV semi-amplitude of the primary star, $i$ is the orbital inclination, $P$ is the orbital period, $e$ is the orbital eccentricity, and $G$ is Newton's gravitational constant. 
The left-hand side of Equation~\ref{eq:fm} contains three unknowns $M_1$, $M_2$, and $i$, and so two more constraints are required to pin down these parameters. The amplitude and shape (asymmetry between the superior and inferior conjunctions) of the EVs or tidal RV signals, if detected, provide two more constraints that involves the mass ratio $M_2/M_1$, primary's radius divided by the orbital semi-major axis $R_1/a$, and $i$ \citep[e.g.,][]{1993ApJ...419..344M}, but this introduces yet another free parameter $R_1$. Thus, there are now four unknowns given three constraints, and additional constraint is needed to determine $M_1$, $M_2$, $i$, and $R_1$.
There are several possible approaches to obtain this last information: (i) If the primary's surface brightness and distance are known and the secondary's flux is negligible, the spectral energy distribution (SED) directly constrains $R_1$. (ii) If it is reasonable to assume that the primary is filling its Roche lobe, this assumption provides a relation that involves the masses, $P$, and $R_1$. (iii) If the primary's rotation is tidally synchronized with the orbit, projected rotation velocity of the primary, $v_1\sin i=(2\pi R_1/P)\,\sin i$, also provides the necessary information. 

In this paper, we use $v_1\sin i$, which is largely determined by the width of the absorption lines and so is presumably more robust against the secondary source of light. We derive this information from a high-resolution spectrum of the primary star.
If the secondary flux turns out to be non-negligible, $R_1$ from (i), as well as the EV amplitude, will be biased; see the next section for an example. The assumption in (ii) may not be justified in X-ray faint systems \edit1{including detached binaries}.

\subsection{V723 Mon as a Test Case}\label{ssec:intro_v723mon}

V723 Monocerotis (HD 45762) has been known as an SB1 in which the primary giant star shows EVs. \citet{\jaya} combined the RVs and EVs with the primary radius $R_1=24.0\pm0.9\,R_\odot$ from SED modeling to determine the component masses and orbital inclination as $M_1=1.00\pm0.07\,M_\odot$, $M_2=3.04\pm0.06\,M_\odot$, and $i=87.0^{+1.7}_{-1.4}\,\mathrm{deg}$. The nearly edge-on orbit should cause the companion to be eclipsed by the primary giant, but no such signals (as expected for a bright, stellar companion) were detected; thus, they argued that the unseen companion is likely a black hole.
\cite{\masuda} used tidal RVs instead of the photometric EVs to measure the masses adopting the above $R_1$, and obtained consistent results ($M_1=0.82\pm0.14\,M_\odot$, $M_2=2.95\pm0.17\,M_\odot$, and $i>80\,\mathrm{deg}$ as 68\% limits).

Later on, the black hole scenario was refuted by \cite{\eb}, who applied a spectral disentangling technique to a time series of high-resolution spectra from Keck/HIRES and identified the secondary set of lines from the G-subgiant secondary that are broadened and shallow due to its rapid rotation. They also showed that the observed SED is better explained by considering the secondary's flux. Both models consistently suggest that the secondary's bolometric luminosity is as large as $\approx 70\,\%$ of that of the primary, implying that the SED-based $R_1$ and the EV amplitude adopted by \citet{\jaya} are significantly overestimated and underestimated, respectively. Taking the secondary's light into account, \cite{\eb} revised the masses and inclination to be: $\moneeb$, $\mtwoeb$, and $\inceb$.\footnote{We note that this estimate is also based on the assumption that the secondary star is {\it not} eclipsed by the primary giant, without which the uncertainties will be larger; see Section~\ref{sec:discussion} for more details.} The total mass is smaller, and the mass ratio is higher than estimated in \citet{2021MNRAS.504.2577J} and \citet{\masuda}.
This example illustrates the importance of considering the secondary source of flux --- whose presence may be {\it a priori} unknown --- for accurate mass measurements. 
This could remain to be an issue even for genuine SB1s involving a compact object, as the emission from the accretion flow could play a similar role as the light from the secondary star; this was the argument of \citet{2012ApJ...757...36K}. 

The above cited analysis of tidal RVs by \cite{\masuda} did not rely on photometric EVs and was not affected by the underestimation of its amplitude. However, the analysis adopted the (overestimated) stellar radius determined from the SED analysis that was biased by the secondary flux and thus was not free from this systematics.\footnote{Another source of discrepancy was that \cite{\masuda} imposed a prior lower limit of $0.5\,M_\odot$ for the primary giant, which as we saw above turned out to be larger than its more plausible value.}
In this paper, we replace the SED-based constraint on $R_1$ with that on $v_1\sin i$, so that the inference does not rely on absolute flux information that is systematically affected by the secondary,
and perform a mass measurement using tidal RVs {\it without explicitly taking into account the secondary's flux}.
The flux contamination from the secondary is significant, but it resembles a continuum: the spectral lines are broad and shallow \citep{\eb}, and can only be identified with a careful analysis with multiple high-resolution spectra. Thus, the system serves as a \edit1{good} test case to simulate how the spectrum-based mass measurement works in real SB1s with flux contamination, such as X-ray systems.

\subsection{Modeling of Tidal RVs}\label{ssec:intro_tidalrv}

The tidal RV signal itself has been recognized for a long time as a source of spurious orbital eccentricities in spectroscopic binaries \citep{1941PNAS...27..168S}, and has been modeled
as the flux-weighted mean of the velocity field on the stellar surface 
\citep{1941PNAS...27..168S, 1959cbs..book.....K, 1976ApJ...203..182W, 2000A&A...364..265O}.
\citet{\masuda}, however, pointed out that such an implementation is inadequate for quantitative modeling of this signal detected with high signal-to-noise: the RVs derived from distorted spectral lines do not necessarily agree with the flux-weighted mean of the stellar surface velocity, in a manner sensitive to how exactly the RV values are extracted from the distorted spectral lines.
In this paper, we use the model of \citet{\masuda} that explicitly takes into account this difference for a mass measurement solely based on spectra.

\section{Measurement of $v\sin i$ using Subaru/IRD Spectrum}\label{sec:spectrum}

As discussed in Section~\ref{ssec:intro_principle}, the RV data including tidal effects needs to be supplemented with $v\sin i$ of the primary star to determine the binary mass. We perform our own measurement of $v\sin i$ of V723 Mon using a high-resolution, near-infrared spectrum obtained with Subaru/IRD. The result will be used as prior information for the RV modeling in Section~\ref{sec:tidalrv}.

\subsection{Observation and Data Reduction}\label{ssec:spectrum_data}

We observed V723 Mon (HD~45762) using the InfraRed Doppler instrument \citep[IRD,][]{2012SPIE.8446E..1TT, 2018SPIE10702E..11K} on the 8.2~m Subaru telescope on UT 2023 March 6 (program ID: S23A-029, PI: K.~Masuda), and obtained a single $R\sim70,000$ spectrum in $Y$, $J$, and $H$-bands (0.97--1.75~$\mu$m). The exposure time was set to 200~seconds, and the signal-to-noise of $\sim 150$ per pixel was achieved at $1.55\,\mu\mathrm{m}$.
The spectrum was reduced using the public pipeline {\tt PyIRD} \citep{pyIRD},\footnote{\url{https://github.com/prvjapan/pyird}} which performs the aperture extraction and tracing, hot pixel masking, removal of two-dimensional pattern noise of the HAWAII-2RG detector, and wavelength calibration using a ThAr lamp in a fully automated manner.

\begin{figure*}[htbp]
    \centering
    \epsscale{1.12}
    \plotone{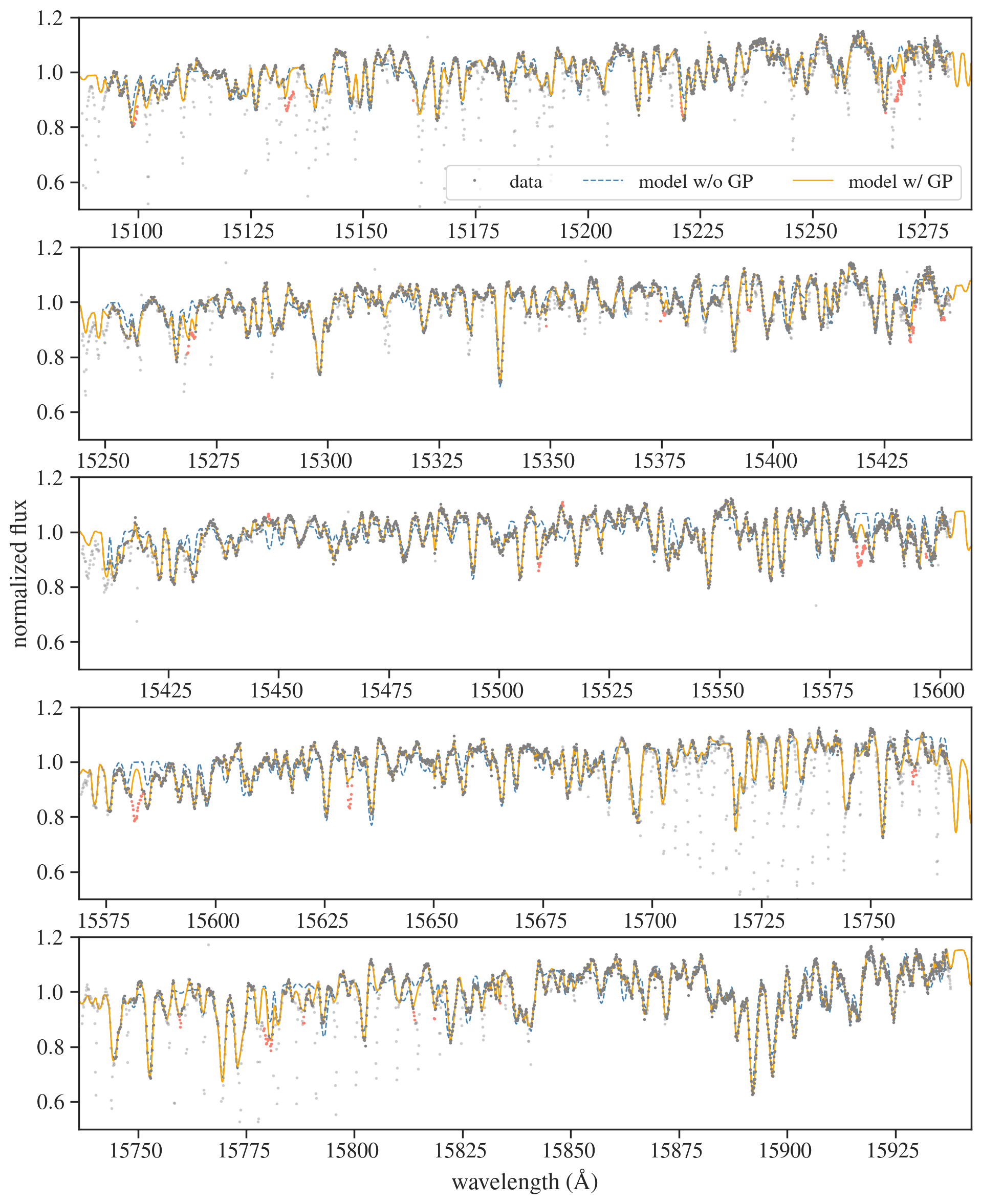}
    \caption{Subaru/IRD spectrum of V723 Mon. The gray dots show the reduced data, and those with lighter color corresponds to the masked pixels (Section~\ref{ssec:spectrum_data}). 
    The red dots show the points removed as outliers during iterative fitting (Section~\ref{ssec:spectrum_model}).
    The blue dotted lines show a physical spectrum model, computed as the mean of the posterior models. The orange solid lines show the mean prediction of a model including a Gaussian process noise model; see Section~\ref{ssec:spectrum_model} for details. 
    }
    \label{fig:spectrum}
\end{figure*}

The resulting 1D spectrum was further processed as following, order by order:
\begin{enumerate}
\item Masking remaining outliers: Although {\tt PyIRD} performed hot pixel masking, we identified additional outliers in the target spectrum that were also seen in the flat spectrum. 
We applied a median filter with a window size of 67 data points to the flat spectrum, and masked the pixels whose flux values deviated more than 10\% away from the median filtered spectrum.
\item Masking telluric lines: We modeled the sky transmittance using SKYCAL Sky Model Calculator\footnote{\url{https://www.eso.org/observing/etc/bin/gen/form?INS.MODE=swspectr+INS.NAME=SKYCALC}} and masked the pixels where the transmittance is less than 98\%.
\item Blaze normalization: We applied a median filter with a window size of 117 data points to the target spectrum from step 2, and divided the original target spectrum by this median-filtered spectrum. The normalization of the spectrum was set so that the mean value of the median-filtered spectrum is unity.
\end{enumerate}
The processed spectrum is shown in Figure~\ref{fig:spectrum} with gray dots, where the flux values in the masked pixels are shown with \edit1{lighter} colors. Considering the computational cost of spectrum fitting, we decided to model the five orders in $H$-band shown here, which exhibit highest signal-to-noise ratios and modest impact of telluric contamination.

\subsection{Spectrum Modeling}\label{ssec:spectrum_model}

\begin{deluxetable}{l@{\hspace{.1cm}}cc@{\hspace{.1cm}}c}[htbp]
\tablecaption{Results of IRD Spectrum Modeling.}\label{tab:spectrum}
\tablehead{
\colhead{} & \colhead{\textbf{w/ dilution}} & \colhead{w/o dilution} & \colhead{prior} 
}
\startdata
$T_{\rm eff}$ (K) & $3802_{-30}^{+26}$ & $3894_{-17}^{+18}$ & $\unif(3500, 7000)$\\
$\log g$ (cgs) & $1.08_{-0.08}^{+0.04}$ & $1.06_{-0.06}^{+0.03}$ & $\unif(1, 5)$\\
$\mathrm{[M/H]}$ & $-0.33_{-0.05}^{+0.05}$ & $-0.57_{-0.03}^{+0.03}$ & $\unif(-1, 0.5)$\\
$\mathrm{[\alpha/Fe]}$ & $0.01_{-0.01}^{+0.00}$ & $0.01_{-0.01}^{+0.00}$ & $\unif(0, 0.4)$\\
$v\sin i$ (km/s) & $15.7_{-0.4}^{+0.3}$ & $15.8_{-0.5}^{+0.3}$ & $\unif(0, 31.5)$\\
$\zeta$ (km/s) & $2.1_{-0.9}^{+1.2}$ & $3.3_{-0.8}^{+0.8}$ & $\unif(0,10)$\\
dilution $D$ & $0.19_{-0.03}^{+0.03}$ & $0$ (fixed) & $\unif(0, 1)$ or fixed\\
$q_1=(u_1+u_2)^2$ & $0.16_{-0.16}^{+0.13}$ & $0.16_{-0.16}^{+0.15}$ & $\unif(0,1)$\\
$q_2={u_1\over 2(u_1+u_2)}$ & $0.21_{-0.21}^{+0.15}$ & $0.19_{-0.19}^{+0.12}$ & $\unif(0,1)$\\
\textit{(GP parameters)}\\
$\ln \rho$ & $-3.45_{-0.02}^{+0.02}$ & $-3.44_{-0.02}^{+0.02}$ & $\unif(-5,0)$\\
$\ln l$ (\AA) & $-0.45_{-0.03}^{+0.03}$ & $-0.45_{-0.03}^{+0.03}$ & $\unif(-5,2)$\\
$\ln s$ & $-5.33_{-0.04}^{+0.04}$ & $-5.33_{-0.04}^{+0.05}$ & $\unif(-10,-3)$\\
\textit{(order 15100\AA--)}\\
normalization & $1.057_{-0.004}^{+0.003}$ & $1.055_{-0.003}^{+0.003}$ & $\unif(0.8,1.2)$\\
slope & $0.12_{-0.01}^{+0.01}$ & $0.12_{-0.01}^{+0.01}$ & $\unif(-0.2,0.2)$\\
RV (km/s) & $-35.3_{-0.3}^{+0.3}$ & $-35.3_{-0.3}^{+0.3}$ & $\unif(-50.0, -18.6)$\\
\textit{(order 15250\AA--)}\\
normalization & $1.049_{-0.004}^{+0.003}$ & $1.048_{-0.003}^{+0.003}$ & $\unif(0.8,1.2)$\\
slope & $0.09_{-0.01}^{+0.01}$ & $0.10_{-0.01}^{+0.01}$ & $\unif(-0.2,0.2)$\\
RV (km/s) & $-35.5_{-0.4}^{+0.3}$ & $-35.5_{-0.4}^{+0.3}$ & $\unif(-50.0, -18.6)$\\
\textit{(order 15425\AA--)}\\
normalization & $1.045_{-0.003}^{+0.003}$ & $1.045_{-0.003}^{+0.003}$ & $\unif(0.8,1.2)$\\
slope & $0.07_{-0.01}^{+0.01}$ & $0.08_{-0.01}^{+0.01}$ & $\unif(-0.2,0.2)$\\
RV (km/s) & $-36.0_{-0.3}^{+0.3}$ & $-36.0_{-0.2}^{+0.3}$ &  $\unif(-50.0, -18.6)$\\
\textit{(order 15575\AA--)}\\
normalization & $1.053_{-0.004}^{+0.003}$ & $1.053_{-0.004}^{+0.003}$ & $\unif(0.8,1.2)$\\
slope & $0.11_{-0.01}^{+0.01}$ & $0.11_{-0.01}^{+0.01}$ & $\unif(-0.2,0.2)$\\
RV (km/s) & $-35.3_{-0.3}^{+0.3}$ & $-35.3_{-0.3}^{+0.3}$ & $\unif(-50.0, -18.6)$\\
\textit{(order 15750\AA--)}\\
normalization & $1.083_{-0.004}^{+0.004}$ & $1.084_{-0.004}^{+0.004}$ & $\unif(0.8,1.2)$\\
slope & $0.17_{-0.01}^{+0.01}$ & $0.16_{-0.01}^{+0.01}$ & $\unif(-0.2,0.2)$\\
RV (km/s) & $-36.2_{-0.2}^{+0.2}$ & $-36.2_{-0.2}^{+0.3}$ & $\unif(-50.0, -18.6)$\\
\hline
\enddata
\tablecomments{Values listed here report the medians and $68\%$ highest posterior density intervals of the marginal posteriors. $\mathcal{U} (a,b)$ denotes the uniform  probability density function between $a$ and $b$.
}
\end{deluxetable}

We interpolate a grid of synthetic spectra, broaden the lines considering macro-turbulence, $v\sin i$, and instrumental resolution, and fit the observed spectrum directly to derive all the relevant parameters including $v\sin i$.

\subsubsection{The Model}

We used {\tt iSpec} \citep{2014A&A...569A.111B, 2019MNRAS.486.2075B} to compute synthetic spectra ($R\sim 250,000$) over a grid of effective temperature $T_{\rm eff}$ (3500--7500~K at intervals of 250~K), surface gravity $\log g$ (1--5 at intervals of 1), metallicity [M/H] ($-1$--0.5 at intervals of 0.25), and $\alpha$ enrichment [$\alpha$/Fe] of 0 and 0.4. The microturbulence velocity was computed deterministically from $T_{\rm eff}$, $\log g$, and [M/H] using an empirical relation. 
The synthesis with {\tt iSpec} was performed using {\tt turbospectrum} \citep{1998A&A...330.1109A, 2012ascl.soft05004P}, {\tt MARCS} model atmospheres \citep{2008A&A...486..951G}, atomic line lists from VALD \citep{2011BaltA..20..503K}, and the solar abundance from \citet{2007SSRv..130..105G}.
We linearly interpolated the grid of continuum-normalized synthetic spectra to obtain the raw spectrum model $f_{\rm raw}(\lambda; \teff, \logg, \mh, \afe)$, where $\lambda$ denotes the wavelength.

The raw spectrum was then convolved with the broadening kernel $M(\lambda)$ that takes into account macrotuburlence, rigid rotation, limb-darkening, and instrumental broadening as described in \citet{2011ApJ...742...69H}. For the macroturbulence velocity field, the radial--tangential model \citep{2005oasp.book.....G} with a single velocity parameter $\zeta$ was adopted. The rigid rotation is parameterized by the line-of-sight component of equatorial rotation velocity $\vsini$. We adopted the quadratic limb-darkening law with two coefficients, $u_1$ and $u_2$. The instrumental broadening was assumed to be Gaussian corresponding to $R=70,000$. The broadened spectrum was then Doppler shifted according to the star's radial velocity (RV) $v$. Thus, our physical model is
\begin{align}
\notag
    f_{\rm phys}(\lambda) = &M(\lambda; v\sin i, \zeta, u_1, u_2, v) \\
\label{eq:fphys}
    &\quad \star f_{\rm raw}(\lambda; \teff, \logg, \mh, \afe)
\end{align}
\edit1{In reality, the line profile of the tidally deformed primary varies with the orbital phase, which is not fully described by the single kernel $M$ as defined above. The systematic error introduced by this model incompleteness will be evaluated at the end of Section~\ref{ssec:spectrum_model} and incorporated in the tidal RV analysis in Section~\ref{sec:tidalrv}.}

To model the observed spectrum, we also fit the normalization and slope of the spectrum in each order, as well as dilution from sources other than the primary star. Namely,
\begin{align}
\label{eq:fmodel}
    f_{\rm model}(\lambda) = f_{\rm base}(\lambda)\,
    \left[ (1-D)\,f_{\rm phys}(\lambda) + D \right]
\end{align}
and
\begin{align}
    f_{\rm base}(\lambda; c_0, c_1) = c_0 + c_1 {{\lambda - \lambda_{\rm mean}} \over {\lambda_{\rm max}-\lambda_{\rm min}}},
\end{align}
where $\lambda_{\rm mean}, \lambda_{\rm max}, \lambda_{\rm min}$ are the mean, maximum, and minimum values of $\lambda$ in each order, respectively. \edit1{The fractional dilution} $D$ is assumed to be independent of the wavelength.
As discussed in Section~\ref{ssec:intro_v723mon}, we already know that $D$ is non-zero due to the secondary's flux, but here we do not use this prior knowledge to simulate a situation where such information is lacking.

\subsubsection{The Inference}

We compute $f_{\rm model}$ order-by-order, setting the parameters $c_0$, $c_1$, and $v$ separately in each order.\footnote{We fit order-dependent RVs to take into account the small offsets in the wavelength calibration.} 
For each order labeled by $k$, we adopt a multivariate-normal likelihood:
\begin{equation}
    \ln{\mathcal{L}^{(k)}} = -\frac{1}{2} (f_{\rm obs}^{(k)} - f_{\rm model}^{(k)})^T \Sigma^{-1} (f_{\rm obs}^{(k)} - f_{\rm model}^{(k)}) -\frac{1}{2} \ln{|2 \pi \Sigma|}.
\end{equation}
Here $f_{\rm obs}$ is the observed flux, and the flux covariance between the $i$th and $j$th pixels is modeled using the Mat\'ern-3/2 covariance function and the white-noise term:
\begin{align}
    \Sigma_{ij} &= (\sigma_i^2+s^2) \delta_{ij} + \nonumber \\
      & \rho^2 \left( 1 + \frac{\sqrt{3}|\lambda_i - \lambda_j|}{l} \right) \exp\left(-\frac{\sqrt{3}|\lambda_i -  \lambda_j|}{l} \right),
\end{align}
where $\sigma_i$ is the flux error calculated assuming Poisson noise, $s$ accounts for the white noise component in excess of $\sigma_i$ if any, $\delta_{ij}$ is the Kronecker delta, $\rho$ represents the amplitude of the covariance, and $l$ represents the scale length of the covariance.
The log-likelihood for the entire data spanning five orders $\ln\mathcal{L}$ was then computed as $\sum_k \ln\mathcal{L}^{(k)}$.

We first minimized $-\ln\mathcal{L}$ to derive an optimal set of parameters. Because we identified clear data--model mismatches presumably due to missing lines in the model, we iteratively removed $3\sigma$ outliers from the likelihood calculation until none is found; they are marked with red dots in Figure~\ref{fig:spectrum}. Then we sampled the parameter sets from the posterior distribution $\propto \mathcal{L} \times \pi(\theta)$ using the No-U-Turn Sampler \citep[NUTS,][]{DUANE1987216, 2017arXiv170102434B} as implemented in {\tt NumPyro} \citep{bingham2018pyro, phan2019composable}, where the prior $\pi(\theta)$ is assumed to be separable for each parameter and is summarized in Table~\ref{tab:spectrum}. We ran four chains for 1,500 steps after which the split Gelman--Rubin statistic $\hat{R}$ was $<1.01$ for each parameter \citep{BB13945229}. The results are summarized in Table~\ref{tab:spectrum} and the models are shown in Figure~\ref{fig:spectrum}. 

The resulting parameters are broadly consistent with those based on disentangled optical spectra \citep[][Table~1]{\eb}. This makes sense given that the absorption lines of the secondary star are broad and shallow.
Interestingly, the inferred value of the dilution factor $D=0.19\pm0.03$ is also comparable to the prediction of \cite{\eb} based on the two-star fit to the SED and spectra (see their Figure~11, bottom); here $D$ defined in Equation~\ref{eq:fmodel} corresponds to $f_2/f_{\rm tot}$ in \citet{\eb}. 

We also repeated a similar analysis fixing $D=0$ (third column in Table~\ref{tab:spectrum}). While the inferred metallicity is decreased, as expected, $v\sin i$ turns out to be insensitive to the choice of $D$. Again this makes sense considering that this parameter is primarily determined by the widths of the lines rather than their depths. Thus the impact of dilution is minimal: even if we do not know {\it a priori} that the secondary object is bright and erroneously assume that $D=0$, the result is not biased as long as $v\sin i$ is large enough to be measurable, yet not too large so that individual lines are still well resolved.

We fit the posterior samples for $v\sin i$ and $\zeta$ with truncated normal distributions with the lower limits of zero, and obtained the \edit1{mean and standard deviation} of $15.7$ and $0.4$ for $\vsini$, and $2.0$ and $1.1$ for $\zeta$, respectively. 
These estimates do not take into account the systematic errors due to the fact that the line profile model adopted here assumes a spherical star and is not fully adequate for the tidally deformed primary star, as was mentioned below Equation~\ref{eq:fphys}. Strictly speaking, this effect will cause the inferred $\vsini$ to be dependent on the orbital phase, as was noted in previous works \citep{2014Obs...134..109G, 2021MNRAS.504.2577J}. 
\edit1{To estimate the systematic error due to this model incompleteness, we simulated a time series of the phase-dependent, asymmetric line profiles using the model in Section~\ref{sec:tidalrv}, and fitted them using the model in Section~\ref{ssec:spectrum_model} that ignores the effects of tidal deformation. Based on comparison between the $\vsini$ values used to simulate the former realistic line profiles and those recovered from the latter simplified model, we estimated the systematic error to be $\sim 1\,\kms$, which is larger than the statistical error ($0.4\,\mathrm{km/s}$) as obtained above. Thus, we adopt $\vsini = 15.7 \pm 1.0\,\kms$ in the following analysis.}
This estimate agrees with $\vsini=15\pm2\,\kms$ based on the disentangled HIRES spectra \citep[optical, $R=60,000$;][]{\eb}, $\vsini=16\pm2\,\kms$ based on the SES spectra \citep[optical, $R=55,000$;][]{2012AN....333..663S}, and $\vsini\sim15\,\kms$ by \citet{2014Obs...134..109G}, and is smaller than $\vsini=17.9\pm0.4\,\kms$ by \citet{2021MNRAS.504.2577J} based on the optical, $R\approx 220,000$ Potsdam Echelle Polarimetric and Spectroscopic Instrument \citep[PEPSI;][]{2015AN....336..324S} spectra by $\sim 2\sigma$. 
We suspect that the slight tension with the last estimate is due to the contamination of the secondary's spectra that is more significant in the optical than in the near-infrared wavelengths \citep{\eb}.

\section{Modeling of Tidal RVs}\label{sec:tidalrv}

We model the RV variations of the primary giant both due to orbital motion and tidal deformation, along with the constraint on $v\sin i$ from Section~\ref{sec:spectrum}, to measure the binary mass without using absolute flux information.

\begin{figure*}[htbp]
     \centering
     \epsscale{1.15}
     \plotone{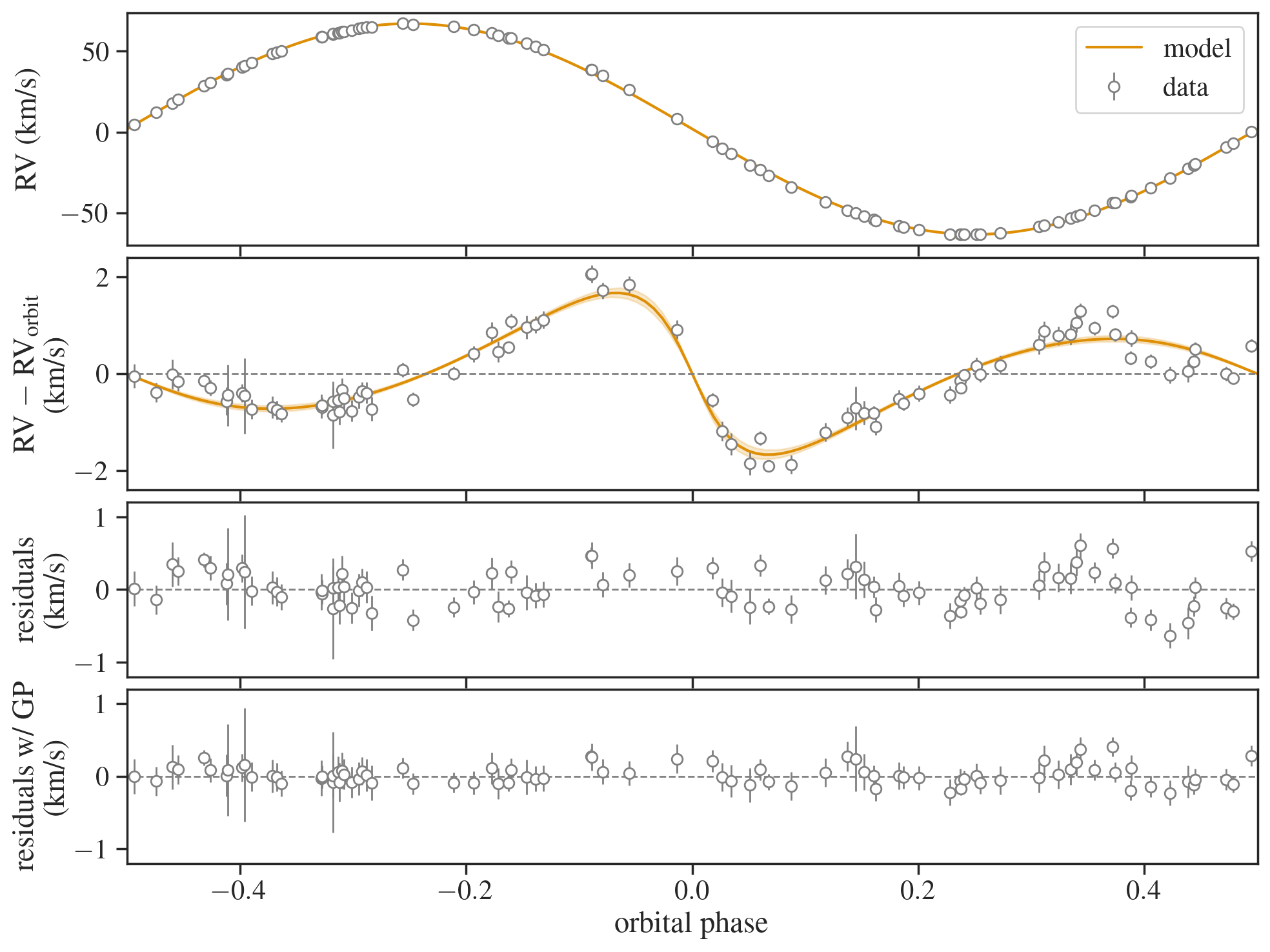}
     \caption{
     The observed and modeled RVs as a function of orbital phase. The white filled circles are the RV data from \citet{2012AN....333..663S}. The orange solid line and the shaded region respectively show the mean and standard deviation of the physical models computed for posterior samples of the parameters. 
    {\it (Top)} --- Raw RVs.
    {\it (Upper-middle)} --- RVs relative to the orbital component plus $\gamma$ ($v_{\rm tidal}$ in Equation~\ref{eq:vmodel}).
    {\it (Lower-middle)} --- Residuals relative to the full physical model ($v_{\rm tidal}$ in Equation~\ref{eq:vmodel}).
    {\it (Bottom)} --- RVs relative to the full physical model + mean prediction of the Gaussian process noise model.
     } 
     \label{fig:rv}
\end{figure*}

\subsection{The Data and the Prior}

We modeled the same set of RV data from \citet{2012AN....333..663S} as analyzed in \citet{\masuda}. The velocities were derived from high-resolution ($R=55,000$), optical (338--882~nm) spectra from the STELLA \'{e}chelle spectrograph (SES) on the 1.2~m STELLA-I telescope at the Teide Observatory \citep{2004AN....325..527S, 2008SPIE.7019E..0LW, 2010AdAst2010E..19S}, via an order-by-order cross correlation analysis using a synthetic spectrum \citep{1993sssp.book.....K} that roughly matches the target spectral classification as a template.
We also adopt $v\sin i =15.7\pm1.0\,\kms$ and $\zeta=2.0\pm1.1\,\kms$ based on the Subaru/IRD spectrum analyzed in Section~\ref{sec:spectrum} as the prior information.

\subsection{The Model}

We adopt the same RV model as developed by \citet{\masuda},\footnote{The code is available on the author's GitHub: \url{https://github.com/kemasuda/rochev}}
in which the Roche model is adopted to describe the surface of the giant star and the RV variations due to tidal deformation were evaluated following the cross-correlation procedure as was actually used to derive RV values by \citet{2012AN....333..663S}. The main procedures are summarized as follows: 
\begin{enumerate}
    \item The distributions of the flux and line-of-sight velocity over the surface of the giant star is computed at the times of observations, given the time of inferior conjunction\footnote{Here defined as the conjunction where the secondary subgiant (less luminous but more more massive) is in front of the primary giant.} $t_0$, orbital period $P$, inclination $i$, 
    primary's mass $M_1$, companion's mass $M_2$,
    primary's radius $R_1$, gravity-darkening coefficient $y$, and two limb-darkening coefficients $u_1$ and $u_2$ for the quadratic law.
    The flux and velocity values are evaluated at 768 {\tt HEALPix/healpy} pixels \citep{2005ApJ...622..759G, Zonca2019}. 
    The line-of-sight velocity of each pixel is computed assuming that the rotation of the giant star is synchronized with the orbital motion. 
    The gravity- and limb-darkening coefficients are constrained to be around the values determined as a function of a single effective wavelength $\lambda_{\rm eff}$ using the relation in \citet{2011A&A...529A..75C}.
    The radius $R_1$ of the tidally deformed star is defined to be the value at the points on the stellar equator perpendicular to the star--companion axis.
    \item The flux and velocity distributions are used to compute the absorption line profile of the giant star, given 
    the macroturbulence velocity $\zeta$. 
    For the macrotubulence velocity field, we adopt a radial--tangential model \citep{2005oasp.book.....G} and assume the common velocity dispersions for both directions.
    \item The line profile model was used to simulate the cross-correlation \edit1{function (CCF) as results from cross-correlating the observed spectrum with a theoretical template spectrum,} whose line profile is assumed to be a Gaussian with the standard deviation (in the velocity space) of $\beta$. The tidal RV, $v_{\rm tidal}$, is computed as the velocity corresponding to the CCF peak.
    \item The center-of-mass velocities $v_0$ of the giant star are computed as 
    \begin{equation}
    \label{eq:vmodel_kepler}
        v_0(t) = - K\,\sin\left[2\pi(t-t_0)\over P\right] + \gamma,
    \end{equation}
    where $K$ is computed from the above $M_1$, $M_2$, $i$, and $P$, and $\gamma$ is the RV zero point. Then the full physical model for the observed RVs is
    \begin{align}
    \label{eq:vmodel}
        v_{\rm model}(t) = v_0(t) + v_{\rm tidal}(t).
    \end{align}
\end{enumerate} 
Thus the physical RV model is specified by the set of parameters $\theta=\{M_1, M_2, R_1, t_0, P, i, \zeta,\lambda_{\rm eff}, \beta, \gamma\}$.

In \citet{\masuda}, we adopted the prior from the SED analysis for the primary's radius $R_1$, but here we instead use $v_1\sin i=(2\pi R_1/P)\sin i$ derived from high-resolution spectra (Section~\ref{sec:spectrum}). Given $v_1\sin i$ and $i$, $R_1$ is given deterministically. 
To improve the sampling efficiency, we also use
\edit1{$A = a/a_{\rm crit}$}, where
\begin{align}
    {R_1 \over a_{\rm crit}} = {0.49q^{2/3} \over {0.6q^{2/3}+\ln(1+q^{1/3})}}, \quad q={M_1 \over M_2}
\end{align}
\citep{1983ApJ...268..368E}.
Given $q$, this equation converts \edit1{$A$ into $a/R_1=A a_{\rm crit}/R_1$}. Combined with $R_1$ above, $a/R_1$ then fixes $a$, and so the total mass via Kepler's third law. So with $q$, the masses of both stars are specified.
In summary, we adopt a set of ($v\sin i$, $i$, $q$, \edit1{$A$}) instead of $(M_1, M_2, R_1, i)$ as the parameters directly sampled from the prior.

For the noise model, \citet{\masuda} adopted independent and identical Gaussian distributions for each data point. In this work, we adopt the following multivariate normal distribution $\mathcal{N}$ to take into account possible correlations over time in the noise:
\begin{align}
    p(v_{\mathrm{obs}}|\theta, \rho, l, s)
    = \mathcal{N}(v_\mathrm{obs}; v_{\rm model}(t; \theta), \Sigma(\rho,l,s)),
\end{align}
where the covariance matrix $\Sigma$ is given by a Mat\'ern-3/2 kernel and a white noise term:
\begin{align}
\notag
    \Sigma_{ij} &= (\sigma_i^2+s^2) \delta_{ij} +\\
      & \rho^2 \left( 1 + \frac{\sqrt{3}|t_i - t_j|}{l} \right) \exp\left(-\frac{\sqrt{3}|t_i - t_j|}{l} \right),
\end{align}
where $t_i$ and $\sigma_i$ denote the time and assigned error of the $i$th RV measurement. The parameters $\rho$ and $l$ correspond to the amplitude and length of the correlation, respectively, and $s$ takes into account the scatter in excess of the assigned errors, if any.

\subsection{The Inference}

\begin{deluxetable}{l@{\hspace{.1cm}}c@{\hspace{.15cm}}c}[!ht]
\tablecaption{Results of Tidal RV Modeling.}\label{tab:rv}
\tablehead{
\colhead{parameter} & \colhead{value} & \colhead{prior} 
}
\startdata
$M_1$ ($M_\odot$) & $0.46_{-0.09}^{+0.12}$ & \nodata \\
$M_2$ ($M_\odot$) & $2.5_{-0.2}^{+0.2}$ & \nodata \\
$M_2/M_1$ & $5.5_{-0.9}^{+1.1}$ & $\lunif(1,10)$\\
$R_1$ ($R_\odot$) & $19.7_{-1.2}^{+1.2}$ & \nodata \\
$a/R_1$ & $4.7_{-0.2}^{+0.2}$ & \nodata \\
$A=a/a_{\rm Roche}$  & $1.15_{-0.03}^{+0.04}$ & $\lunif(1,5)$\\
$v\sin i$ (km/s) & $16.4_{-1.0}^{+1.0}$ & $\mathcal{N}(15.7, 1.0, 0)$\\
$\zeta$ (km/s) & $2.2_{-1.1}^{+1.1}$ & $\mathcal{N}(2.0, 1.1, 0)$\\
$\cos i$ & $0.13_{-0.09}^{+0.12}$ & $\unif(0,1)$\\
$K$ (km/s) & $65.31_{-0.06}^{+0.05}$ & \nodata \\
$\gamma$ (km/s)  & $1.88_{-0.04}^{+0.04}$ & $\unif(-10, 10)$\\
$t_0$ (day)\tablenotemark{$\dagger$}& $764.9675_{-0.0094}^{+0.0091}$  & $\unif(764.5, 765.5)$\\
$P$ (day) & $59.9371_{-0.0014}^{+0.0013}$ & $\unif(59,9, 60)$\\
$\beta$ (km/s) & $3.2_{-0.4}^{+0.5}$ & $\mathcal{N}(3, 0.5, 2.5)$\\
$\lambda_\mathrm{eff}$ (nm) & $602_{-98}^{+110}$ & $\mathcal{N}(635.0, 123.5, 38)$\\
$u_1$ & $0.68_{-0.17}^{+0.21}$ & $\mathcal{N}(u_1(\lambda_\mathrm{eff}), 0.1)$\\
$u_2$ & $0.15_{-0.15}^{+0.12}$ & $\mathcal{N}(u_2(\lambda_\mathrm{eff}), 0.1)$\\
$y$ & $0.52_{-0.09}^{+0.11}$ & $\mathcal{N}(y(\lambda_\mathrm{eff}), 0.1)$\\
$\ln \rho$ (km/s) & $-1.6_{-0.2}^{+0.2}$ & $\unif(-5,1)$\\
$\ln l$ (day) & $0.9_{-0.4}^{+0.3}$ & $\unif(-3, 5)$\\
$\ln s$ (km/s) & $-3.9_{-0.9}^{+0.9}$ & $\unif(-5,1)$\\
\hline
\enddata
\tablenotetext{\dagger}{Time with respect to the first measurement, $\mathrm{BJD}=2454065.57926$.}
\tablecomments{Values listed here report the medians and 16th/84th percentiles of the marginal posteriors. $\mathcal{U} (a,b)$ denotes the uniform  probability density function between $a$ and $b$. $\mathcal{LU} (a,b)$ denotes the log-uniform probability density function between $a$ and $b$. $\mathcal{N} (a,b,c)$ denotes the Gaussian probability density function with the mean of $a$ and the scale of $b$, truncated at the lower bound $c$ when it is given.
}
\end{deluxetable}

We again used NUTS to sample from the joint posterior distribution:
\begin{align}
    p(\theta, \rho, l, s|v_{\mathrm{obs}}) \propto p(v_{\mathrm{obs}}|\theta, \rho, l, s)\,\pi(\theta, \rho, l, s),
\end{align}
where the prior probability density function $\pi$ was assumed to be separable for each parameter, as summarized in Table~\ref{tab:rv}. 
The adopted priors are mostly the same as in \citet{\masuda} (see their Section~2.4), except that we explicitly defined the prior PDFs for $v\sin i$ and $A=a/a_{\rm Roche}$ instead of $M_1$ and $R_1$. 
We ran two chains for 3,000 steps after which the split Gelman--Rubin statistic $\hat{R}$ was estimated to be $<1.02$ for all parameters \citep{BB13945229}. The results are summarized in Table~\ref{tab:rv} and the models are shown in Figure~\ref{fig:rv}.

\section{Discussion}\label{sec:discussion}

\begin{figure*}[htbp]
     \centering
     \epsscale{1.15}
     \plotone{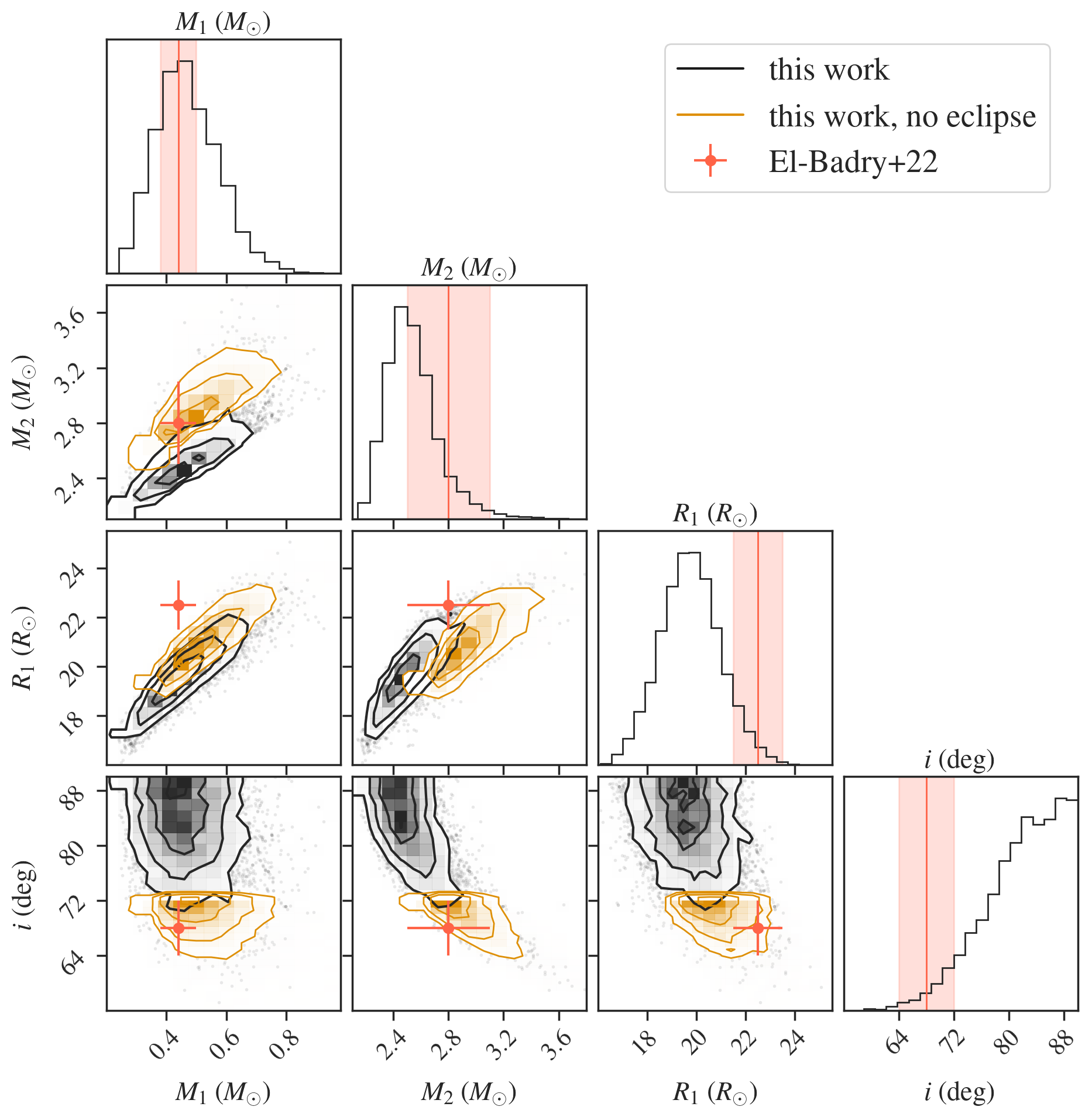}
     \caption{Corner plot \citep{corner} for the posterior samples obtained from tidal RV modeling in Section \ref{sec:tidalrv}. The black contours and histograms correspond to our fiducial analysis in Section~\ref{sec:tidalrv}. The red crosses are the values derived by \citet{\eb} using EVs, $R_1$ from SED, and the constraint that the secondary is not eclipsed by the primary. 
     The orange contours show the subsample from our tidal RV modeling that satisfies $i<72\,\mathrm{deg}$, which corresponds to the last constraint adopted by \citet{\eb}.}  
     \label{fig:corner}
\end{figure*}

We successfully determined the binary parameters using spectroscopic information (i.e., RVs and line widths) alone, and found $M_1=\mone$ and $M_2=\mtwo$. Although we did not explicitly take into account the significant contamination of the secondary's flux, our results show a reasonable agreement with $M_1=\moneeb$ and $M_2=\mtwoeb$ obtained from the absolute flux and EVs, after carefully modeling the secondary's contribution \citep{\eb}.
This serves as a proof of concept that modeling of tidal RVs is useful as an alternative means to measure the mass of tidally deformed SB1s.

As shown in the corner plot for $M_1$, $M_2$, $R_1$, and $i$ (Figure~\ref{fig:corner}), however, our modeling (black contours) prefers $\sim 2\sigma$ larger $i$ and smaller $R_1$ than those derived by \citet{\eb} (red crosses) so that the resulting $v\sin i$ remains similar.
This may point to a bias in the inclination estimate in our tidal RV modeling. 
One possible explanation, which is specific to V723 Mon, is the bias in measured RVs due to the secondary's spectral lines. Although it is difficult to fully consider its impact when analyzing only RV values, we discuss its partial impact below in Section~\ref{ssec:discussion_secondary} and find that the tension is slightly relieved by considering its effects.
Another possibility is that our model for analytic evaluation of CCFs may be too simplified in that it assumes a common shape for all the absorption lines.
This type of systematic errors may be reduced by directly modeling the line shapes rather than RV values calculated from the distorted lines, as has also been performed in the analyses of the Rossiter-McLaughlin effects for transiting exoplanets \citep[e.g.,][]{2022PASP..134h2001A}.

That said, we note that the tension is at least partly due to the additional constraint adopted by \citet{\eb} that the secondary subgiant is not eclipsed by the primary giant: they inferred $i\gtrsim 62\,\mathrm{deg}$ based on the photometric EV amplitude, and then excluded $i>72\,\mathrm{deg}$ based on the lack of eclipses. The former lower limit alone in fact agrees well with our constraint from tidal RVs \edit1{($i>63\,\mathrm{deg}$ as the 99.7\% lower limit; see also Figure~\ref{fig:corner}). If we also incorporate the latter constraint $i<72\,\mathrm{deg}$ in our analysis,} we do find a better agreement with \citet{\eb}, as shown by orange contours in Figure~\ref{fig:corner}.
Thus, we conclude that the tidal RV modeling is at least as useful as modeling of EVs, although it may also suffer from its own systematics.

\subsection{Impact of Secondary Lines on Measured RVs}\label{ssec:discussion_secondary}

Strictly speaking, the secondary flux in V723 Mon is not merely continuous light.
\citet{\eb} pointed out that the secondary's absorption lines --- even though they are broad and shallow --- can well affect the measured RVs of the primary giant. \citet{\eb} simulated that the RV bias induced by the secondary as a function of orbital phase is sinusoidal to the zeroth order and can cause the RV semi-amplitude $K$ to be underestimated by $\approx 1\,\kms$. We also confirmed this with our simulations using the binary parameters in \citet{\eb} and the line profile model as described in Section~\ref{sec:tidalrv}.\footnote{
For certain values of the assumed RV semi-amplitude and $\vsini$ of the secondary subgiant, we also observed a sudden drop in the RV bias around the quadrature phases. This happened when the RV separation became larger than the line widths of the subgiant. We assume this did not happen in the actual data because we do not see such structured RV residuals in Figure~\ref{fig:rv}.
}

To check on the impact of this bias in $K$, we performed a modified version of the analysis in Section~\ref{sec:tidalrv} assuming that $K$ in Equation~\ref{eq:vmodel_kepler} is smaller by $1\,\kms$ than the value computed from the other physical parameters. 
From this analysis, we obtained $M_1=0.47_{-0.09}^{+0.12}\,M_\odot$, $M_2=2.61_{-0.15}^{+0.20}\,M_\odot$, $R_1=19.83_{-1.1}^{+1.2}\,R_\odot$, $\cos i=0.12_{-0.08}^{+0.12}$. These values remain consistent with those in Table~\ref{tab:rv}, though $M_2$ and $R_2$ become slightly closer to the values given by \citet{\eb}. \edit1{This modified model did not significantly improve the goodneess of the fit.}


\section{Summary and Conclusion}\label{sec:summary}

We determined the physical parameters of the binary system V723 Mon by modeling RV variations of the tidally deformed primary star along with its $v\sin i$ derived from Subaru/IRD near-infrared spectra, where the former includes the orbital motion signal and apparent variations due to the deformation of absorption lines. 
To our knowledge, this is the first quantitative mass measurement of an SB1 system using spectroscopic information alone without relying on stellar models nor on the absolute flux information. Despite that we did not explicitly model the secondary's flux that is significant in this system, the resulting masses ($M_1=\mone$ and $M_2=\mtwo$) show a reasonable agreement with those determined from modeling of ellipsoidal variations ($M_1=\moneeb$ and $M_2=\mtwoeb$) and the SED that explicitly takes into account the secondary's flux \citep{\eb}.
This demonstrates that modeling of tidal RV signal serves as a useful alternative means for mass measurements of tidally deformed SB1s that is more robust against contaminating light sources other than the primary star \citep{\masuda}.
The method is potentially useful for secure mass measurements of systems of particular interest, including dormant compact object binaries on tighter orbits than being identified from the Gaia survey data \citep[e.g.,][]{2023MNRAS.518.1057E,2023ApJ...946...79T}.
\edit1{More generally, this method can also be beneficial for systems where the photometric ellipsoidal variation signal is significantly affected by non-primary light sources, such as a tertiary star, foreground or background stars, or accretion flow from mass transfer.}
The comparison between the results based on photometric EVs and those based on tidal RVs also suggests a possible bias in the orbital inclination derived from tidal RV modeling whose origin is not yet understood. 
We expect that this possible systematics can be further tested and potentially be reduced by modeling the orbital-phase-dependent distortion of line profiles directly rather than by modeling the apparent RV changes due to the distortion. This is beyond the scope of the present paper. 




\acknowledgements

The authors thank the anonymous reviewer for very careful reading and for providing detailed and thoughtful comments, which greatly improved the quality and clarity of the manuscript.
This research is based in part on data collected at the Subaru Telescope, which is operated by the National Astronomical Observatory of Japan. We are honored and grateful for the opportunity of observing the Universe from Maunakea, which has the cultural, historical, and natural significance in Hawaii.
KM acknowledges support by JSPS KAKENHI grant No.~21H04998.
M.T. is supported by JSPS KAKENHI grant No.~24H00242.

\facility{Subaru (IRD)}

\software{corner \citep{corner}, HEALPix \citep{2005ApJ...622..759G}, healpy \citep{Zonca2019}, JAX \citep{jax2018github}, NumPyro \citep{bingham2018pyro, phan2019composable}, PyIRD \citep{pyIRD}
}




\bibliography{references_masuda}
\bibliographystyle{aasjournal}



\end{document}